\documentclass[epj]{MYsvjour}
\usepackage{graphics}
\begin{document}
\title{Challenges for tau physics at the TeraZ}
\author{Antonio Pich
}                     
%
%
\institute{IFIC, Universitat de València -- CSIC, Catedrático José Beltrán 2, E-46980 Paterna, Spain.}
\date{Received: date / Revised version: date}
%
\abstract{
The very high statistics, low backgrounds and clean back-to-back kinematics of a TeraZ facility would provide an optimal laboratory for precision measurements of the $\tau$ properties. A few important topics in $\tau$ physics where very relevant contributions could be made are highlighted.
\PACS{
      {14.60.Fg}{tau properties}   \and
      {13.35.Dx}{tau decays}\and
      {12.15.-y}{electroweak interactions}\and
      {12.38.-t}{QCD} 
     } 
} 
\maketitle
\section{Introduction}
\label{intro}

The precise investigation of the third fermion family can provide crucial hints on our understanding of the flavour problem. The heavier fermions are expected to be more sensitive to the unknown dynamics responsible for the observed structure of three sequential generations, exhibiting a very broad range of different mass scales and mixing angles. The leptonic nature of the $\tau$ provides, in addition, a clean environment to perform accurate tests of the electroweak interaction. Moreover, since the $\tau$ is heavy enough to decay into hadrons, it constitutes an ideal tool for studying low-energy effects of QCD \cite{Pich:2013lsa}.

Our knowledge of the $\tau$ lepton was spectacularly boosted by the clean  data samples collected by the LEP experiments, at the $Z$ peak, which promoted $\tau$ physics to the level of precision tests. Since then, the B factories have accumulated a much larger statistics of about $10^9$ $\tau^+\tau^-$ pairs, establishing strong upper bounds of few times $10^{-8}$ 
on many neutrinoless lepton-flavour-violating $\tau$-decay rates \cite{Bevan:2014iga}. The on-going Belle-II experiment will significantly increase the available samples to around $4.6\times 10^{10}$ produced $\tau^+\tau^-$ pairs \cite{Kou:2018nap}, which could 
push these limits further down by more than one order of magnitude. 

However, the experimental values of many $\tau$ properties, such as the main branching ratios or the inclusive spectral distributions, are still dominated by the LEP data, in spite of the much lower statistics that was available at that $e^+e^-$ collider. The most complete data sample provided by the ALEPH experiment contains only $3.3\times 10^5$ reconstructed $\tau$ decays \cite{Schael:2005am}, but nevertheless, it plays still a leading role in many phenomenological analyses and determines the current limits on the theoretical interpretation of the most interesting observables. The special kinematical configuration of the back-to-back $\tau^+\tau^-$ pairs emerging from a $Z$ decaying at rest offers an ideal scenario for $\tau$ analyses, with controllable systematic uncertainties and relatively low backgrounds. Tagging one $\tau$ and selecting the event in the other hemisphere, one can collect a very large set of $\tau$ decays without any bias or selection requirement, which allows for a precise measurement of the $\tau$ branching fractions without any need for an external normalization. Obviously, inclusive studies of $\tau$ decay distributions benefit also from this clear kinematic advantage.

The TeraZ option of a future FCC-ee collider running at the $Z$ peak would produce an enormous data sample of $1.7\times 10^{11}$  $\tau^+\tau^-$ pairs \cite{Abada:2019lih}, in extremely clean kinematic and background conditions, opening a broad range of interesting opportunities. In the next sections, I survey a few examples of relevant tests of our currently accepted theoretical framework that could be significantly improved at a TeraZ facility.

\section{Lepton universality}
\label{sec:universality}

In the Standard Model (SM), all leptons couple to the $W^\pm$ bosons with exactly the same strength: $g_e = g_\mu = g_\tau \equiv g$. The most accurate phenomenological tests of the universality of the leptonic charged-current couplings are summarized in Table~\ref{tab:ccuniv}. The strongest constraints come from $\pi$ and $\tau$ decays and confirm the flavour universality with a $0.15$\% precision. The accuracy of the $\tau$ determination of $|g_\mu/g_e|$ is directly set by the current uncertainty on the experimental ratio $\Gamma_{\tau\to\mu}/\Gamma_{\tau\to e}$, which could be improved at TeraZ. Improvements on $|g_\tau/g_e|$ would require also more precise measurements of the $\tau$ lifetime and mass. The lifetime could be accurately measured either at Belle-II or at the $Z$ peak, while a future tau-charm factory operating at the $\tau^+\tau^-$ threshold \cite{Barniakov:2019zhx,Luo:2019xqt} could certainly determine $m_\tau$ with high precision.

\begin{table}[tb]\centering
\caption{Experimental determinations of the ratios \ $g_\ell/g_{\ell'}$ \cite{Pich:2013lsa,Aad:2020ayz,Zyla:2020zbs}.}
\label{tab:ccuniv}
\begin{tabular}{llllll}
\hline\noalign{\smallskip} 
& $\Gamma_{\tau\to\mu}/\Gamma_{\tau\to e}$ &
 $\Gamma_{\pi\to\mu} /\Gamma_{\pi\to e}$ &
 $\Gamma_{K\to\mu} /\Gamma_{K\to e}$ &
 $\Gamma_{K\to\pi\mu} /\Gamma_{K\to\pi e}$ &
 $\Gamma_{W\to\mu} /\Gamma_{W\to e}$
\\ \noalign{\smallskip} \hline\noalign{\smallskip}
 $|g_\mu/g_e|$
 & $1.0017\; (16)$ & $1.0010\; (9)$ & $0.9978\; (18)$ & $1.0010\; (25)$ & $0.998\; (4)$
\\ \noalign{\smallskip} \hline\hline\noalign{\smallskip}
& $\Gamma_{\tau\to e}/\Gamma_{\mu\to e}$ &
 $\Gamma_{\tau\to\pi}/\Gamma_{\pi\to\mu}$ &
 $\Gamma_{\tau\to K}/\Gamma_{K\to\mu}$ &
 $\Gamma_{W\to\tau}/\Gamma_{W\to\mu}$
\\ \noalign{\smallskip} \hline\noalign{\smallskip}
 $|g_\tau/g_\mu|$
 & $1.0011\; (14)$ & $0.9965\; (26)$ & $0.986\; (7)$ & $1.004\; (16)$
\\ \noalign{\smallskip} \hline\hline\noalign{\smallskip}
& $\Gamma_{\tau\to\mu}/\Gamma_{\mu\to e}$
 & $\Gamma_{W\to\tau}/\Gamma_{W\to e}$
\\ \noalign{\smallskip} \hline\noalign{\smallskip}
 $|g_\tau/g_e|$
 & $1.0028\; (15)$ & $1.022\; (12)$
\\ \noalign{\smallskip} \hline
\end{tabular}
\end{table}

The LEP measurements of $\Gamma_{W\to\tau}/\Gamma_{W\to e,\mu}$ suggested $2.4\sigma$ and $2.6\sigma$ deviations from lepton universality in $|g_\tau/g_e|$ and $|g_\tau/g_\mu|$, respectively, at the 1\% level, which were very difficult to reconcile with the indirect constraints from $\tau$ decay at the 0.15\% level \cite{Pich:2013lsa}.  The recent ATLAS determination of $\Gamma_{W\to\tau}/\Gamma_{W\to \mu}$ \cite{Aad:2020ayz}, in perfect agreement with the SM expectation, has eliminated this long-standing anomaly, showing the importance of performing new precise measurements of statistically-limited observables. The ATLAS measurement alone would imply $|g_\tau/g_\mu| =0.996\pm 0.007$. The much larger error in the table reflects the sizeable discrepancy with the old LEP value. A preliminary CMS measurement of the $W$ leptonic branching fractions \cite{CMS:2021qxj} fully confirms the ATLAS result.

The universality of the leptonic couplings of the $Z$ boson was precisely tested at LEP \cite{ALEPH:2005ab,ALEPH:2010aa}. The measured ratios of vector and axial-vector couplings \cite{Pich:2013lsa}
\begin{equation}
\label{eq:NCuniversality}
\frac{v_\mu}{v_e}\, =\, 0.961\pm 0.061\, ,
\qquad
\frac{v_\tau}{v_e}\, =\, 0.959\pm 0.029\, ,
\qquad
\frac{a_\mu}{a_e}\, =\, 1.0002\pm 0.0013\, ,
\qquad
\frac{a_\tau}{a_e}\, =\, 1.0019\pm 0.0015\, ,
\end{equation}
are in perfect agreement with the SM. A more precise determination would obviously be possible at the TeraZ, together with significantly improved precision tests of the electroweak interaction.

\section{Lorentz structure of the tau decay amplitude}
\label{sec:Lorentz}

The leptonic decay amplitude $\ell^-\to\nu_\ell\ell'^-\bar\nu_{\ell'}$
can be parametrised in a model-independent way by writing the most general, local, derivative-free, lepton-number conserving, four-lepton interaction Hamiltonian, consistent with locality and Lorentz invariance. It contains 
ten operators with their corresponding complex couplings $g^n_{\epsilon\omega}$ \cite{Pich:2013lsa}:
\begin{equation}
\label{eq:hamiltonian}
{\cal H} \; =\;  4\, \frac{G_{\ell'\ell}}{\sqrt{2}}\;
\sum_{n,\epsilon,\omega}\,
g^n_{\epsilon\omega}\,
\left[ \overline{\ell'_\epsilon}
\Gamma^n {(\nu_{\ell'})}_\sigma \right]\,
\left[ \overline{({\nu_\ell})_\lambda} \Gamma_n
	\ell_\omega \right]\, ,
\end{equation}
The subindices $\omega$ and $\epsilon$ indicate the chiralities (left or right) of 
$\ell$ and $\ell'$, respectively, while $n=S,V,T$ labels the type of interaction (scalar, vector, tensor). Taking out the global normalization factor $G_{\ell'\ell}$ that is determined by the total decay rate, these couplings are bounded to the ranges $|g^S_{\epsilon\omega}|\le 2$, $|g^V_{\epsilon\omega}|\le 1$ and $|g^T_{\epsilon\omega}|\le 1/\sqrt{3}$. In the SM, $g^V_{LL}=1$, while all other couplings are identically zero. Any possible contribution from new-physics at higher scales would result in non-zero values for some of these effective couplings. 

The different contributions can be disentangled  by measuring the energy and angular distribution of the final charged lepton, complemented with polarisation information whenever available. This has been successfully achieved in  $\mu$-decay experiments, which have demonstrated that the bulk of the decay amplitude is indeed of the predicted $V-A$ type, $|g^V_{LL}| > 0.960$ (90\% C.L.) \cite{Zyla:2020zbs} (information from the inverse transition $\nu_\mu e^-\to\mu^-\nu_e$ is also needed), and have established upper bounds on all other couplings. 

The model-independent analysis of the $\tau$ decay is more challenging because of its much shorter lifetime. The polarisation of the secondary charged lepton has never been measured and the inverse production cross section $\sigma(\nu_\tau\ell^-\to\tau^-\nu_\ell)$ seems far out of reach. The initial $\tau$ polarisation can be accessed through the correlated distribution of the $\tau^+\tau^-$ pairs produced in $e^+e^-$ annihilation. The LEP experiments took advantage of the $\tau$ polarization provided by the $Z$ decay to extract the available information. The current constraints are shown in Table~\ref{tab:Michel_tau}. Upper bounds exist for the couplings with an  initial right-handed $\tau$ because the data agree with the SM. However, the Lorentz structure of a left-handed decaying $\tau$ remains undetermined. Although significant improvements should be expected from Belle-II~\cite{Kou:2018nap}, the optimal conditions provided by a TeraZ facility would definitely allow for much higher sensitivities.


\begin{table}[tbh]\centering
\caption{95\% CL experimental bounds on the leptonic $\tau$-decay couplings
\cite{Zyla:2020zbs}}
\label{tab:Michel_tau}
\begin{minipage}{0.45\textwidth}
 \begin{tabular}{llll}
 \hline\noalign{\smallskip} 
 \multicolumn{4}{c}{$\tau^-\to e^-\bar\nu_e\nu_\tau$}
 \\ \noalign{\smallskip} \hline\noalign{\smallskip}
 $|g_{RR}^S| < 0.70$ & $|g_{LR}^S| < 0.99$ & $|g_{RL}^S| \leq 2$
 & $|g_{LL}^S| \leq 2$ \\
 $|g_{RR}^V| < 0.17$ & $|g_{LR}^V| < 0.13$ & $|g_{RL}^V| < 0.52$
 & $|g_{LL}^V| \leq 1$ \\
 $|g_{RR}^T| \equiv 0$ & $|g_{LR}^T| < 0.082$ & $|g_{RL}^T| < 0.51$
 & $|g_{LL}^T|\equiv 0$ 
 \\ \noalign{\smallskip} \hline
 \end{tabular}
\end{minipage}
\hskip .75cm
\begin{minipage}{0.45\textwidth}
 \begin{tabular}{llll}
 \hline\noalign{\smallskip} 
 \multicolumn{4}{c}{$\tau^-\to \mu^-\bar\nu_\mu\nu_\tau$}
 \\ \noalign{\smallskip} \hline\noalign{\smallskip}
 $|g_{RR}^S| < 0.72$ & $|g_{LR}^S| < 0.95$ & $|g_{RL}^S| \leq 2$
 & $|g_{LL}^S| \leq 2$ \\
 $|g_{RR}^V| < 0.18$ & $|g_{LR}^V| < 0.12$ & $|g_{RL}^V| < 0.52$
 & $|g_{LL}^V| \leq 1$ \\
 $|g_{RR}^T| \equiv 0$ & $|g_{LR}^T| < 0.079$ & $|g_{RL}^T| < 0.51$
 & $|g_{LL}^T|\equiv 0$ 
 \\ \noalign{\smallskip} \hline
 \end{tabular}
 \end{minipage}
\end{table}

\section{Constraints on violations of lepton flavour and lepton number}

With the large data samples collected at the B factories, sensitivities of a few times $10^{-8}$ (90\% CL) have been reached in many leptonic ($\tau\to\ell\gamma$, $\tau\to\ell'\ell^+\ell^-$) and semileptonic ($\tau\to\ell P^0$, $\tau\to\ell V^0$, $\tau\to\ell P^0 P^0$, $\tau\to\ell P^+P'^-$) lepton-flavour-violating $\tau$ decays~\cite{Amhis:2019ckw}. Thanks to its very clean signature, competitive limits on the decay $\tau\to 3\mu$
have been also set by LHCb. The absence of final neutrinos makes these decay modes easier to identify, and many final states have very low backgrounds. Thus, in first approximation, the sensitivity scales linearly  with the integrated luminosity. Substantially improved limits, reaching at least $10^{-9}$, are then to be expected from Belle-II~\cite{Kou:2018nap}.

Stringent upper bounds in the range ($2.0 - 8.4) \times 10^{-8}$ (90\% CL) have been also set on the lepton-number-violating decay modes $\tau^-\to\ell^+P^-P'^-$, with $\ell= e,\mu$ and $P,P'=\pi,K$ \cite{Miyazaki:2013yaa}. The combined violation of baryon and lepton quantum numbers has been tested in the decay $\tau^-\to\bar p\mu^-\mu^+$, where an upper limit of $1.8\times 10^{-8}$ (90\% CL) has been established \cite{Sahoo:2020rzr}.


The huge data sample that would be accumulated at a TeraZ facility could increase in a drastic way the sensitivity to new-physics scales in these neutrinoless decay modes, perhaps providing the first clear evidence of lepton-flavour violation in $\tau$ decays. The large number of accessible $\tau$ decay modes violating the lepton flavour would provide complementary information, allowing us to unravel the particular type of dynamics underlying any observed experimental signature.

\section{QCD tests with hadronic tau decays}
\label{sec:QCD}

The semileptonic decays $\tau^-\to\nu_\tau H^-$ probe the matrix element of the left-handed quark current between the vacuum and the hadronic state $H^-$. The observed quantum numbers of the final hadrons make possible to disentangle the vector and axial-vector currents, and also separate the Cabibbo allowed ($\bar d u$) and Cabibbo suppressed ($\bar s u$) sectors. Thus, the $\tau$ provides an excellent laboratory to study the dynamics of the QCD Goldstone bosons ($\pi$, $K$, $\eta$) in the resonance region, around 1~GeV.

An exhaustive study of exclusive decays, with well-identified final states, can be performed at Belle-II. However, the accurate determination of the inclusive hadronic distributions requires a much higher control of experimental systematic uncertainties without introducing relative biases (normalizations, acceptances) among the different final states. The special kinematical and background conditions of a TeraZ facility are optimal for this task.

The inclusive distributions directly measure the absorptive parts (spectral functions) of the two-point correlators of the QCD vector and axial-vector currents. The weighted integrals of these spectral functions can be accurately predicted with short-distance operator-product-expansion techniques. The  dominant perturbative contributions are currently known with an impressive four-loop accuracy, {\it i.e.} at ${\cal O}(\alpha_s^4)$, while non-perturbative corrections are suppressed by at least four powers of the $\tau$ mass (the exact power depends on the weight function adopted) \cite{Pich:2020gzz}.

The Cabibbo-allowed component of the $\tau$ hadronic width can be expressed in the form \cite{Braaten:1991qm}
\begin{equation}
R_{\tau,V+A} \equiv \frac{\Gamma(\tau^-\to\nu_\tau + \mathrm{hadrons}\, [\bar d u])}{\Gamma(\tau^-\to\nu_\tau e^-\bar\nu_e)}  \, =\,
N_C\, |V_{ud}|^2 S_{\mathrm{EW}}\,\left\{ 1 + \delta_{\mathrm{P}} + \delta_{\mathrm{NP}}\right\}\, ,
\end{equation}
where $N_C=3$ is the number of QCD colours and $S_{\mathrm{EW}}=1.0201\pm 0.0003$ \cite{Marciano:1988vm,Braaten:1990ef,Erler:2002mv}
accounts for the electroweak radiative corrections. The non-perturbative contribution $\delta_{\mathrm{NP}}$ turns out to be heavily suppressed by a factor $m_\tau^{-6}$. This small correction can be directly determined from the data, through the study of weighted integrals of the hadronic invariant mass distribution with less-suppressed power corrections~\cite{LeDiberder:1992zhd}. The detailed studies performed by the ALEPH \cite{Schael:2005am,Buskulic:1993sv,Barate:1998uf,Davier:2005xq,Davier:2008sk}, CLEO \cite{Coan:1995nk} and OPAL \cite{Ackerstaff:1998yj} collaborations confirmed long time ago that $\delta_{\mathrm{NP}}$ is indeed below 1\%. 

The perturbative contribution $\delta_{\mathrm{P}}$ has been computed to $O(\alpha_s^4)$ \cite{Pich:2013lsa,Baikov:2008jh}, including resummation of large logarithms~\cite{LeDiberder:1992jjr}, and is very sensitive to the strong coupling~\cite{Braaten:1991qm,Narison:1988ni,Braaten:1988ea}. Since $\alpha_s$ is large at the $\tau$ mass scale, $\delta_{\mathrm{P}}$ completely dominates the theoretical prediction, enhancing very significantly $R_{\tau,V+A}$ by about $\sim 20\%$. The unknown higher-order perturbative corrections are the largest source of theoretical uncertainty~\cite{LeDiberder:1992jjr,Pich:2018jiy}. The most recent analyses, performed with updated ALEPH $\tau$ decay distributions, give \cite{Davier:2013sfa,Pich:2016bdg}
\begin{equation}
\alpha_s^{(n_f=3)}(m_\tau^2) =    0.328\pm 0.013 \, ,
\qquad\qquad\longrightarrow\qquad\qquad
\alpha_s^{(n_f=5)}(M_Z^2) = 0.1197\pm 0.0015\, ,
\end{equation}
in nice agreement with the value obtained from the $Z$ hadronic width at $s=M_Z^2$, $\alpha_s^{(n_f=5)}(M_Z^2) = 0.1199\pm 0.0029$ \cite{Zyla:2020zbs}.  The comparison of these two determinations of the strong coupling, at two very different energy scales, constitutes a beautiful test of the predicted QCD running:
\begin{equation}
\left.\alpha_s^{(n_f=5)}(M_Z^2)\right|_{Z}  - \left.\alpha_s^{(n_f=5)}(M_Z^2)\right|_\tau =  0.0002 \pm 0.0029_Z\pm 0.0015_\tau \, .
\end{equation}
Accurate measurements of the spectral hadronic distributions would make possible to pin down more precisely the non-perturbative corrections. A very large data sample is specially needed to access the higher range of kinematically-allowed energies, where the large size of the experimental errors is hampering the precision of current theoretical analyses.

These spectral distributions contain also precious information on the non-trivial structure of the QCD vacuum. The difference of the vector and axial-vector correlation functions is a pure non-perturbative object (perturbation theory gives a null contribution to all orders in $\alpha_s$) that is sensitive to the QCD chiral-symmetry breaking. Current data have already been used~\cite{Pich:2020gzz,Rodriguez-Sanchez:2016jvw,Pich:2021yll} to extract several relevant parameters characterizing the QCD vacuum and low-energy couplings of Chiral Perturbation Theory, the low-energy effective theory of the QCD Golsdstone bosons. 

Useful constraints on new physics beyond the SM can be also extracted from the hadronic $\tau$ decays~\cite{Cirigliano:2018dyk}, making use of model-independent low-energy effective Lagrangians analogous to Eq.~(\ref{eq:hamiltonian}). The semileptonic $\tau$ decays offer also the possibility to perform tests of CP symmetry through appropriate rate asymmetries. BaBar reported a $2.8\sigma$ anomaly in the $\tau^+\to\bar\nu_\tau \pi^+K_S$ CP asymmetry \cite{BABAR:2011aa}, which so far has not been confirmed by Belle \cite{Bischofberger:2011pw}. The significance of this type of tests would greatly benefit from large data samples with controlled systematics and low backgrounds.

\section{Determination of the Cabibbo quark mixing} 
 
The unitarity tests of the quark mixing matrix rely on a broad set of precise experimental measurements and theoretically-determined hadronic form factors and radiative corrections. Recent improvements on the estimated radiative corrections to the neutron and nuclear $\beta$ decays have generated a $3\sigma$ unitarity anomaly when combined with the kaon determinations of $|V_{us}|$ \cite{Zyla:2020zbs}. Independent precise estimates of $|V_{ud}|$ and $|V_{us}|$ are needed to clarify the situation.

A very clean and precise determination of $|V_{us}|$ could be obtained from the
ratio of the inclusive $|\Delta S|=1$ and $|\Delta S|=0$ $\tau$ decay widths (normalised to the electronic width) $R_{\tau,S}/R_{\tau,V+A}$ \cite{Gamiz:2004ar,Gamiz:2002nu}. This experimental ratio directly measures $|V_{us}/V_{ud}|$, in the limit of exact SU(3) symmetry. Taking into account the PDG value of $V_{ud}$ and the small SU(3)-breaking correction $\delta R_{\tau,\mathrm{th}}= 0.240\pm 0.032$ \cite{Gamiz:2006xx,Pich:1999hc,Pich:1998yn}, one finds
\begin{equation}
\label{eq:Vus}
|V_{us}|\, =\, 
\left( \frac{R_{\tau,S}}{\frac{R_{\tau,V+A}}{|V_{ud}|^2} -\delta R_{\tau,\mathrm{th}}}\right)^{1/2}
\, =\,  0.2194\pm 0.0019\, ,
\end{equation}
which is $4.2\sigma$ lower than the unitarity expectation 
$|V_{us}|^{\mathrm{uni}} = \sqrt{1-|V_{ud}|^2-|V_{ub}|^2} = 0.2278\pm 0.0006$.
Note, however, that the only fully-inclusive measurement of the $|\Delta S|=1$ $\tau$ decay distribution was performed at LEP and suffers from a very scarce statistics. Moreover, the exclusive $\tau$ branching ratios measured by BaBar and Belle are on average slightly lower than the LEP ones, a systematic effect that although slowly improving it is not yet well understood \cite{Zyla:2020zbs}.

Clearly, $\tau$ decays could provide a very accurate determination of $|V_{us}|$, which does not involve theoretically-estimated hadronic form factors or decay constants. Unfortunately, the current data on Cabibbo-suppressed $\tau$ decays are not precise enough. A TeraZ facility could obviously resolve the current discrepancies and precisely clarify this important test of the unitarity of the quark mixing matrix.

\section{Tau production in decays of beauty hadrons}

The flavour anomalies recently observed in $b\to c\tau\bar\nu$ and $b\to s\mu^+\mu^-$ decays \cite{Pich:2019pzg}, exhibiting sizeable violations of lepton flavour universality, have triggered a renewed interest in the study of suppressed decays of beauty particles. Whether they just reflect statistical fluctuations and/or underestimated uncertainties, or represent true signals of new phenomena, remains to be understood. Nevertheless, they exhibit the potential sensitivity to physics beyond the SM of precise flavour measurements.

The possible new-physics explanations of these anomalies can be tested through high-statistics analyses of these type of decays in different beauty hadrons (mesons and baryons), and related processes such as $b\to s\tau^+\tau^-$, $B_s\to\tau^+\tau^-$, $B_c\to\tau \nu$ or $D_s\to\tau \nu$. This is another area where an $e^+e^-$ collider running at the $Z$ peak could bring a significantly improved understanding and, perhaps, establish clear evidence of new phenomena.

\section*{Acknowledgements}

This work has been supported in part by the Spanish Government and ERDF funds from the EU Commission [grant FPA2017-84445-P], by the Generalitat Valenciana [grant Prometeo/2017/053], by the EU H2020 research and innovation programme [grant agreement 824093]
and by the COST Action CA16201 PARTICLEFACE.

%
\bibliographystyle{spphys}
\bibliography{TauRefs}
%

\end{document}